\documentclass{osa-article}
\usepackage[per-mode = symbol]{siunitx}
\usepackage{amsmath}
\usepackage{subfig}
\usepackage{multirow}
\usepackage{graphicx}
\usepackage{booktabs}
\usepackage{makecell}
\usepackage{threeparttable}
\usepackage[normalem]{ulem}

\journal{oe}

\articletype{Research Article}

\begin{document}
\title{Multi-channel swept source optical coherence tomography concept based on photonic integrated circuits}
%
\author{Stefan Nevlacsil,\authormark{1,*} Paul Muellner,\authormark{1} Alejandro Maese-Novo,\authormark{1} Moritz Eggeling,\authormark{1} Florian Vogelbacher,\authormark{1} Martin Sagmeister,\authormark{2} Jochen Kraft,\authormark{2} Elisabet Rank,\authormark{3} Wolfgang Drexler,\authormark{3} and Rainer Hainberger\authormark{1}}
\address{\authormark{1}AIT Austrian Institute of Technology GmbH, Giefinggasse 4, 1210, Vienna, Austria\\
\authormark{2}ams AG, Tobelbader Stra\ss e 30, 8141 Premst\"atten, Austria\\
\authormark{3}Center for Medical Physics and Biomedical Engineering, W\"ahringer G\"urtel 18-20, 1090 Vienna, Austria
}
\email{\authormark{*}stefan.nevlacsil@ait.ac.at} 
%
\begin{abstract}
In this paper, we present a novel concept for a multi-channel swept source optical coherence tomography (OCT) system based on photonic integrated circuits (PICs). At the core of this concept is a low-loss polarization dependent path routing approach allowing for lower excess loss compared to previously shown PIC-based OCT systems, facilitating a parallelization of measurement units. As a proof of concept for the low-loss path routing, a silicon nitride PIC-based single-channel swept source OCT system operating at \SI{840}{\nano\metre} was implemented and used to acquire in-vivo tomograms of a human retina. The fabrication of the PIC was done via CMOS-compatible plasma-enhanced chemical vapor deposition to allow future monolithic co-integration with photodiodes and read-out electronics. A performance analysis using results of the implemented photonic building blocks shows a potential tenfold increase of the acquisition speed for a multi-channel system compared to an ideal lossless single-channel system with the same signal-to-noise ratio.
\end{abstract}
%
\section{Introduction}
Optical coherence tomography (OCT) is an imaging technique able to visualize the layer composition in a sample by interfering low-amplitude light reflected from the layer boundaries with a reference light \cite{Drexler.2015, Boer.2017}. The non-invasive nature of OCT makes it well suited for medical applications as well as for industrial inspection. Wider utilization of OCT is, however, limited by the cost and the size of state-of-the-art systems. A reduction of both would enable point-of-care applications with light-weight portable or even handheld systems. This can be achieved by using photonic integrated circuits (PICs) to substitute bulk and/or fiber optics systems which are used for conventional OCT implementations. PICs consist of a combination of photonic building blocks substituting individual fiber-optical and bulk components of an optical setup. In addition, optoelectronic and electronic components can be monolithically co-integrated with PICs allowing for detection and electronic processing of the optical signal on a single chip. Therefore, complete optoelectronic systems can be realized on a single chip within an area of <\SI{1}{cm^2}. Multiple integrated circuits can be fabricated in parallel on a single wafer, which results in a significant cost reduction for each individual system. 

State-of-the-art OCT systems employ the Fourier domain (FD) modality, which offers faster acquisition speeds and higher mechanical stability compared to the time domain modality, typically using a moving reference mirror \cite{Leitgeb.2003}. In FD-OCT the layer information is encoded in the spectral response of the interference signal between sample and reference light. The layer information is extracted via Fourier transformation. The spectral response is obtained in two different ways in so-called spectral domain (SD) and swept source (SS) OCT. While in SD-OCT a broadband light source and a spectrometer is used, in SS-OCT a high-speed wavelength tunable laser source is paired with a synchronized photodiode pair. With respect to PIC-based OCT, the SS-OCT approach holds more promise of high imaging performance because it involves less demanding integrated photonic key building blocks. In particular, SD-OCT requires an integrated spectrometer, e.g.\@ in the form of an arrayed waveguide grating (AWG), which is difficult to implement with sufficiently high quality to compete with state-of-the-art bulk optical systems. In comparison, in SS-OCT the interference signal is directly measured with photodiodes. Several groups have already reported PIC-based SS-OCT systems. Table~\ref{tab:prior_art} summarizes these results and highlights the most important key parameters and implemented integrated components. 

\begin{table}[ht]
\caption{List of previously reported PIC-based SS-OCT systems.}
\centering
\resizebox{\textwidth}{!}{%
\begin{tabular}{ccllll}
\hline
Year & $\lambda_\text{c}$ [nm] & WG technology & Res [\si{\micro\metre}] & Sens [dB] & Integrated components \\ \hline
\cite{Yurtsever.2010} (2010)&1550&\parbox{.18\textwidth}{SOI wire \\ \SI[product-units = power]{220x450}{\nano\metre}}&40&25&\parbox[c][1.3cm][c]{.35\textwidth}{splitter/interferometer (SD) \\ reference arm (18 mm PPL)} \\ \hline
\cite{Yurtsever.2011} (2011)&1550&\parbox{.18\textwidth}{SOI wire \\ \SI[product-units = power]{220x585}{\nano\metre}}&-&-&\parbox[c][1.3cm][c]{.35\textwidth}{splitter/interferometer (SD) \\ reference arm (19.2 mm PPL)} \\ \hline
\cite{Nguyen.2012} (2012)&1310&\parbox{.18\textwidth}{SiN TriPlex wire \\ \SI[product-units = power]{3400x50}{\nano\metre}}&12.7&80&\parbox[c][1.3cm][c]{.35\textwidth}{splitter/interferometer (BD) \\ reference arm} \\ \hline
\cite{Schneider.2014} (2014)&1315&\parbox{.18\textwidth}{SOI \\\SI{220}{\nano \metre}}&100&>40&\parbox[c][1.3cm][c]{.35\textwidth}{splitter/interferometer (SD) \\ reference arm \\ photodiode} \\ \hline
\cite{Yurtsever.2014} (2014)&1310&\parbox{.18\textwidth}{SOI rib \\\SI[product-units = power]{470x220}{\nano\metre} \\ 70 nm etch}&25.5&71*&\parbox[c][1.3cm][c]{.35\textwidth}{splitter/interferometer (BD) \\ reference arm (50.4 cm OPL)} \\ \hline
\cite{Wang.2015} (2015)&1550&-&16&80*& \parbox[c][1.3cm][c]{.35\textwidth}{interferometer (PS, PD, IQ) \\ photodiodes} \\ \hline
\cite{Chang.2016} (2016)&1310&\parbox{.18\textwidth}{SiON wire \\ \SI[product-units = power]{2000x600}{\nano\metre}}&-&67*&\parbox[c][1.3cm][c]{.35\textwidth}{splitter/interferometer (SD) \\ reference arm (end facet) \\ micro ball lens} \\ \hline
\cite{Schneider.2016} (2016)&1315&\parbox{.18\textwidth}{SOI\\\SI{220}{\nano \metre}}&\parbox{.10\textwidth}{ 11 (int) \\ 30 (ext) }&\parbox{.10\textwidth}{53 (int) \\ 64 (ext)}&\parbox[c][1.3cm][c]{.35\textwidth}{splitter/interferometer (BD) \\ reference arm (int) \\ photodiodes} \\ \hline
\cite{Huang.2017} (2017)&1310&SOI&11&86*&\parbox[c][1.3cm][c]{.35\textwidth}{splitter/interferometer (BD)} \\ \hline
\cite{Eggleston.2018} (2018)&1310&-&-&93*&\parbox[c][1.3cm][c]{.35\textwidth}{splitter/interferometer (BD) \\ reference arm \\ photodiodes} \\ \hline
\cite{vanLeeuwen.2018} (2018)&1550&\parbox{.18\textwidth}{SiN TriPlex wire \\ \SI[product-units = power]{3400x50}{\nano\metre}}&15.2&89*&\parbox[c][1.3cm][c]{.35\textwidth}{splitter/interferometer (BD) \\ reference arm (9.2 cm PPL)} \\ \hline

\end{tabular}%

}
\begin{tablenotes}\footnotesize
\item[*] '*' normalized to \SI{1}{\milli\watt} power on the sample
\item[*] $\lambda_\text{C}$~=~central wavelength, WG~=~waveguide, Res~=~axial resolution, Sens~=~sensitivity (SNR with a mirror as sample)
\item[*] SD~=~single detection, BD~=~balanced detection
\item[*] PS~=~polarization sensitive, PD~=~polarization diverse, IQ~=~in-phase and quadrature
\item[*] OPL~=~optical path length, PPL~=~physical path length 
\item[*] int~=~reference path on PIC, ext~=~reference path in free space
\end{tablenotes}
\label{tab:prior_art}
\end{table}

Most of the previously reported PIC-based SS-OCT systems attempted an one-to-one replication of conventional OCT concepts using photonic building blocks. This resulted in a reduced imaging performance compared to state-of-the-art OCT systems because disadvantages of PICs were not compensated by inherent strengths. Although the potential reductions in cost and size are highly beneficial, the practical relevance of PIC-based OCT systems will strongly depend on their imaging performance. The main disadvantages of PIC-based OCT systems are increased excess losses and wavelength dispersion compared to bulk/fiber components. Coupling light into and out of the PIC and on-chip waveguide propagation loss (PL) represent major loss contributions, which affect the achievable signal-to-noise ratio (SNR). As a result of the higher wavelength dispersion, several implementations of the PIC-based SS-OCT systems exhibited a reduction of axial resolution \cite{Yurtsever.2010,Yurtsever.2011,Yurtsever.2014, Schneider.2014, Schneider.2016}.

The inherent strengths of PICs besides cost and size advantages are the capability of combining multiple functions on a monolithic chip and the resulting high mechanical stability of the associated photonic building blocks. For bulk/fiber optics systems multiple functions require additional components and extra efforts for alignment and maintenance. Using photolithography for the PIC fabrication allows a high integration density of photonic building blocks without additional cost, fabrication time, or assembly steps. These advantages become increasingly prevalent with more functionality implemented on the PIC. The high integration density has been exploited to some degree in the implementation of an OCT receiver \cite{Wang.2015} allowing balanced detection, polarization sensitive and/or polarization diversity measurements. Another approach to capitalize on these strengths for OCT is the parallelization of measurements employing multiple light spots emitted from the PIC probing a sample. With multiple measurement spots the time required for a line scan (B-scan) or volumetric scan is significantly reduced. Multiple probing spots have been previously implemented with a cascade of $1\times2$ splitters in a SS-OCT system \cite{Huang.2017}. In this system, each measurement spot has a different on-chip path length resulting in a time division multiplexed interference signal. The interference signal is detected with a single photodiode and the individual imaging positions are separated in post processing. By using only a single photodiode the read-out and processing speed is inherently limited. Moreover, no balanced detection can be used, which causes a high unbalanced relative intensity noise characteristic of SS-OCT \cite{Boer.2017}. Another downside of this implementation is the fact that the cascade of $1\times2$ splitters is also used to recombine the signal in the optical return path, which induces significant excess loss. In conclusion, this approach does not take real advantage of parallelization, which would require a low-loss routing of the sample light towards an interferometer. A dedicated optical return path would further allow the implementation of a Mach-Zehnder configuration and therefore balanced detection, which is a prerequisite for a high-performance SS-OCT system. Complete measurement channels can be formed by co-integrating a photodiode pair for each probing spot, therefore removing the limitation in the read-out speed of a single external photodiode. 

In this paper, we propose a PIC-based multi-channel SS-OCT concept that avoids high routing losses inherent to other PIC-based systems demonstrated so far. At the core of this concept is a low-loss polarization dependent path routing approach, which we experimentally apply in a single-channel configuration for OCT imaging of a healthy human retina. On the basis of these results we provide an in-depth performance analysis of a multi-channel implementation.

\section{Multi-channel SS-OCT concept}
\begin{figure}
\centering
\includegraphics[width=13.3cm]{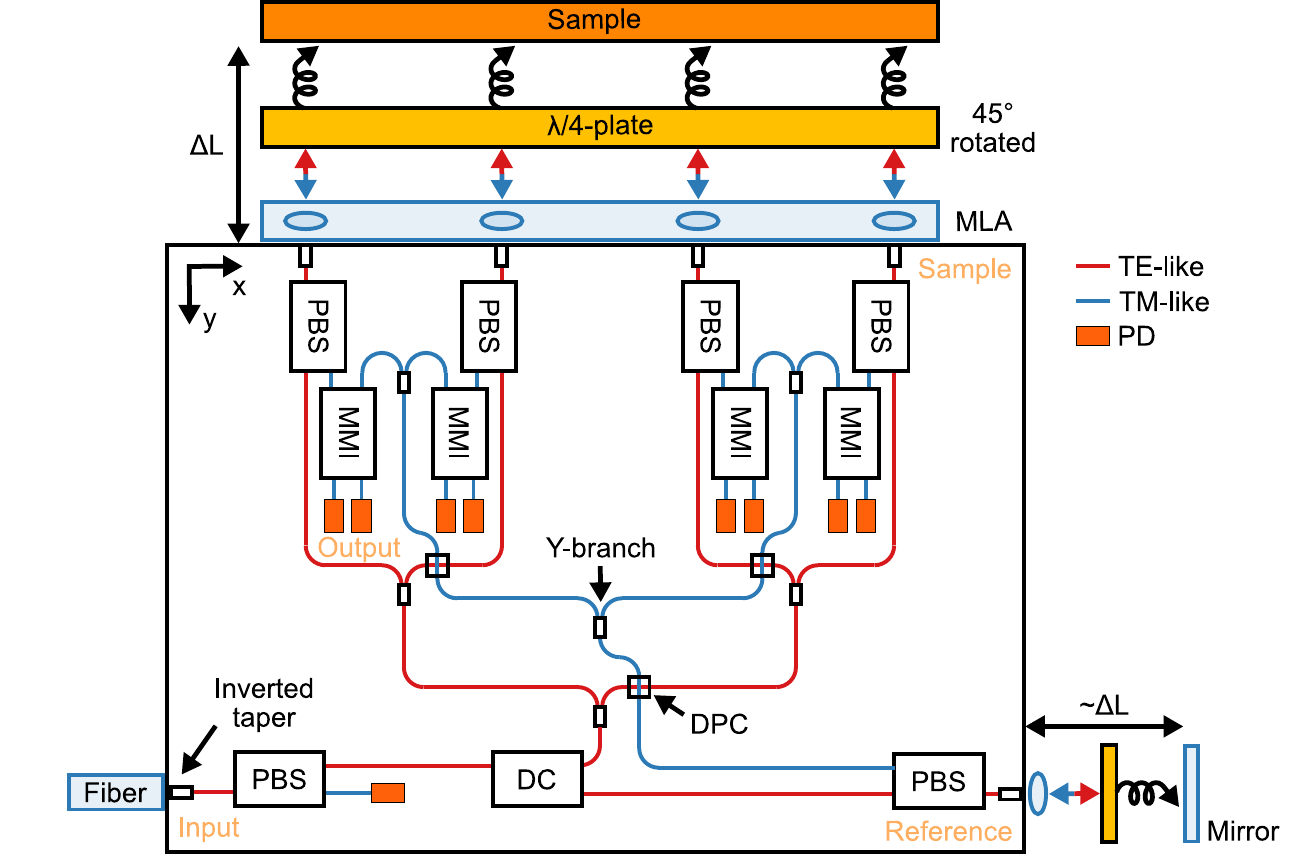}
\caption{Conceptual drawing of a PIC-based four-channel OCT system. The input light is coupled to the PIC with a fiber via an inverted taper. The input polarization is set to transverse electric (TE)-like polarization by minimizing the light measured at the photodiode (PD) at the termination port of the first polarization beam splitter (PBS). The optical power is split into the reference and sample path via a directional coupler (DC) with an asymmetric splitting ratio and into the individual channels with a cascade of Y-branches. On both the sample and the reference path the outgoing light passes a PBS, which redirects the returning light with the orthogonal polarization towards the multimode interferometer (MMI). The polarization rotation from the forward propagating TE-like mode to the backward propagating transverse magnetic (TM)-like mode is achieved outside of the PIC with a $\lambda/4$-plate rotated by 45$^\circ$. On its way from the $\lambda/4$-plate to the sample and reference mirror the light is circularly polarized. The returning reflected light is converted into the orthogonal polarization relative to the outgoing light by passing the $\lambda/4$-plate a second time. The returning reference light crosses the sample path twice, which can be achieved efficiently with low-loss and low-crosstalk dual polarization crossings (DPCs). The highly divergent beams from the PIC are collimated with a microlens array (MLA). The PIC only provides short parts of the sample and reference paths avoiding high propagation losses and wavelength dispersion effects.}
\label{fig:4_channel_PIC}
\end{figure}

Figure~\ref{fig:4_channel_PIC} shows the proposed PIC-based multi-channel OCT concept in a four-channel configuration. In this system, an orthogonal polarization orientation is introduced between light propagating from the source to the sample and light returning from the sample. Therefore, an almost lossless polarization based on-chip path routing is achieved with a polarization beam splitter (PBS). The rotation of the polarization is introduced outside the PIC via a double pass (forward and backward) through a $\lambda/4$-plate which is rotated by 45$^\circ$ relative to the polarization axis of the waveguide. In combination with optoelectronic and electronic components monolithically co-integrated on the PIC a multi-channel SS-OCT system can be realized that holds the promise of meeting the demands for compactness, cost-reduction, and at same time high imaging performance.
The presented concept avoids high routing losses of light by efficiently separating the forward optical path (between the light source and the sample) from the backward optical path (between the sample and the photodiodes). All PIC-based OCT concepts demonstrated so far use an optical power splitter to separate the forward optical path from the backward optical path, which inherently induces significant excess loss associated with the fixed splitting ratio. In fiber optical systems, low-loss optical circulators can be used to separate the forward and backward optical path. However, integrated optical circulators require complex fabrication steps, and even then they exhibit a limited operational wavelength bandwidth in current implementations \cite{Huang.2017b,Pintus.2019}. Broadband PIC-based polarization beam splitters can be implemented without intricate fabrication steps in various waveguide materials \cite{Dai.2018}. Therefore, using a PBS in a polarization dependent path routing allows easy fabrication while avoiding significant loss of source light and of returning light from the sample, which is crucial for a multi-channel system. Another advantage of the proposed routing concept stems from the polarization rotation outside of the PIC. Light reflected before the rotation, e.g.\@ at the end facet, keeps its polarization state and therefore is suppressed on its way towards the interferometer. This reduces reflection artifacts in the OCT measurement, which are caused by similar path lengths of light reflected at the end facets of the sample output and the reference output, as observed in the measurement by Schneider et al. \cite{Schneider.2016}
\section{Retinal OCT imaging at 840\,nm}
The proposed PIC-based multi-channel SS-OCT concept facilitates high imaging speed, while keeping losses to a minimum, and, thus, provides advantages in many fields of application of OCT. In particular, the acquisition speed is of high importance for retinal imaging because of the rapid uncontrollable movement of the eye. Furthermore, OCT angiography is gaining increasing attention for the imaging of microvascular structures in the retina \cite{Boer.2017}. For OCT angiography, multiple tomograms have to be recorded at the same position with a high acquisition rate to reconstruct the blood flow in microvascular structures. Hence, retinal imaging is a good benchmark to evaluate the proposed PIC-based multi-channel OCT concept. With respect to the PIC part of this concept, the polarization based low-loss optical path routing is the most crucial aspect to enable the multi-channel configuration. Hence, in this work we restrict our experiments to the demonstration of the path routing concept for a single-channel SS-OCT system. 

\begin{figure}[ht]
\centering
\includegraphics[width=13.3cm]{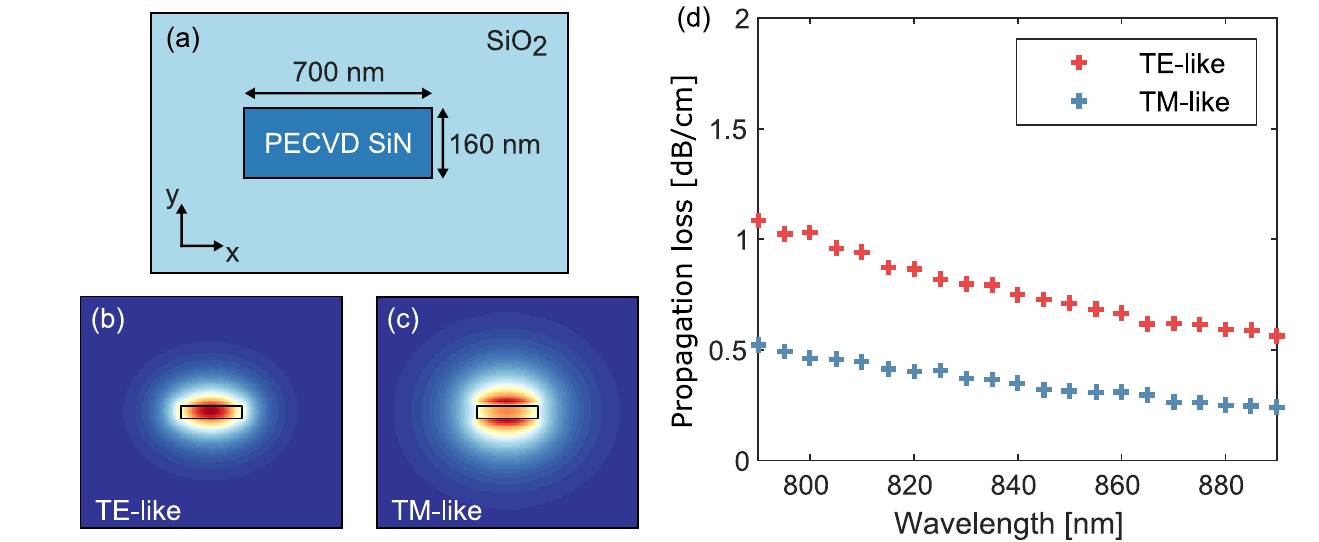}
\caption{(a) Waveguide cross section used for routing on the PIC. Normalized electric field amplitude distribution of light with (b) TE-like and (c) TM-like polarization. (d) Corresponding measured propagation losses in the wavelength range of \SIrange{790}{890}{\nano \metre}.}
\label{fig:prop_loss}
\end{figure}

Since retinal OCT imaging is usually performed in the wavelength region around \SI{840}{\nano \metre}, we use silicon nitride (SiN) waveguides with silicon dioxide (SiO$_2$) as cladding material. Wire waveguides were chosen because of their high optical confinement and versatile application range. A SiN layer thickness of \SI{160}{\nano \metre} was chosen as a compromise between several aspects. \textcolor{black}{One of the main aspects was the coupling efficiency from fiber/bulk optics to the PIC and vice versa. Due to the use of two polarization states at the sample port the coupling has to be optimized for both transverse electric (TE)-like and transverse magnetic (TM)-like polarized light. This requires the use of a quadratic cross section at the end facet of the PIC resulting in similar field distributions for both polarizations. According to simulation and confirmed via measurements, a cross section of \SI{160x160}{\nano\metre} provides a numerical aperture (NA) of about 0.13, which is compatible with a commercially available high NA fiber (Corning RC PM 850) and is still manageable with a microlens. While a smaller cross section would further expand the mode field diameter and allow the use of conventional single mode fibers, a waveguide width of \SI{160}{\nano\metre} is already at the limit of the lithographic resolution. Besides, for a thinner waveguide core the coupling losses to the silicon substrate would become larger because the lower cladding thickness is limited to a given maximum value due to boundary conditions given by the fabrication process. A waveguide width of \SI{700}{\nano \metre} \textcolor{black}{was selected} for routing \textcolor{black}{to ensure single mode propagation for both polarizations.}
}
\textcolor{black}{With a cross section of \SI{700x160}{\nano\metre} the minimum bend radius was selected as \SI{150}{\micro \metre} resulting in bend losses below 0.05 dB per 180$^\circ$ circular bend for both polarizations \cite{Hainberger.02.02.2019}. This radius allows for a multi-channel configuration with a reasonable pitch of \SI{500}{\micro \metre} between two adjacent sample ports.} Figures~\ref{fig:prop_loss}(a-c) show the routing waveguide cross section and the corresponding mode fields.

The SiN waveguides were fabricated at ams AG in a CMOS-compatible process using low-temperature plasma-enhanced chemical vapor deposition (PECVD), deep UV photolithography, and reactive ion etching, which allows monolithic co-integration in the back-end-of-line (BEOL) with silicon photodiodes and read-out electronics \cite{Sagmeister.2018}. The upper and lower SiO$_2$ cladding thickness was set to \SI{3}{\micro \metre}. The chip end facets were deep etched to enable efficient coupling of light from fibers or bulk optics to the waveguides. The refractive indices were determined with ellipsometry to be 1.916 for SiN and 1.455 for SiO$_2$ at \SI{840}{\nano \metre}. The fabricated PICs exhibit propagation losses below \SI{1}{dB/cm} and \SI{0.5}{dB/cm} for TE-like and TM-like polarization, respectively, in the relevant wavelength range, as can be seen in Fig.~\ref{fig:prop_loss}(d). 

\begin{figure}[ht]
\centering
\includegraphics[width=13.3cm]{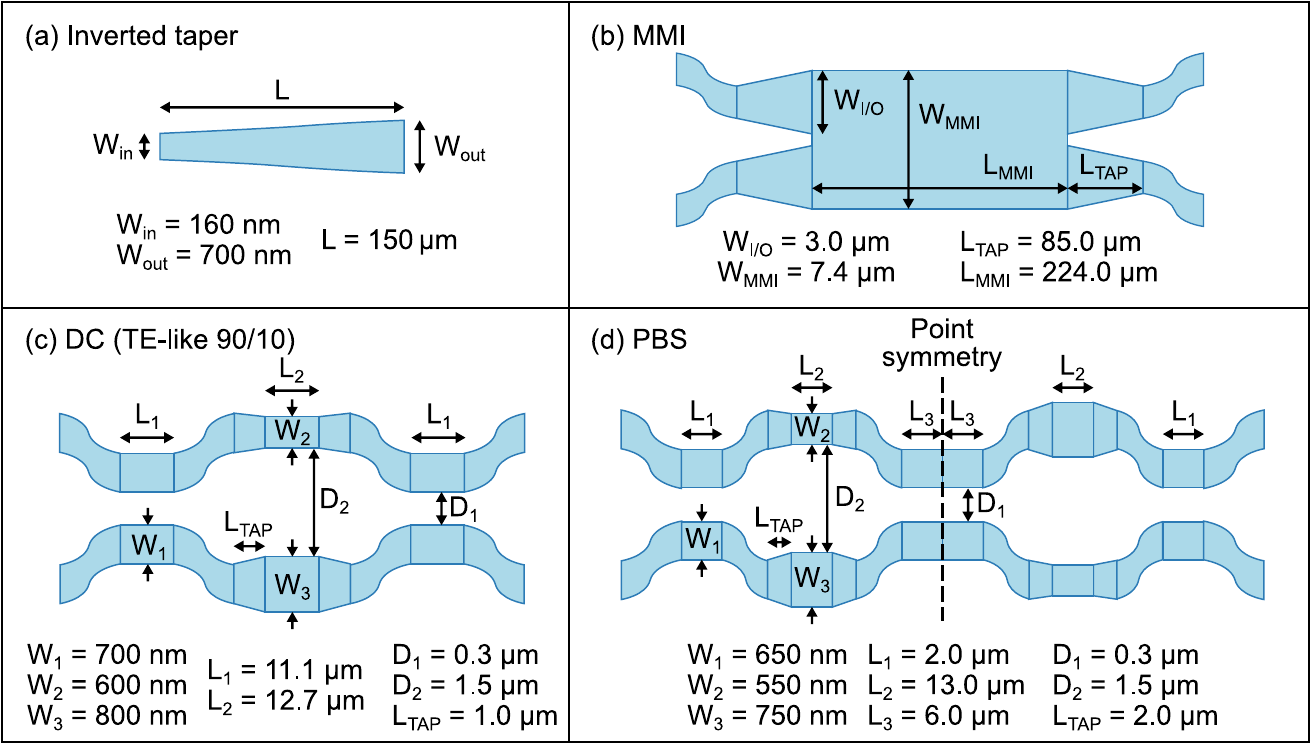}
\caption{Schematics of the broadband photonic building blocks required for low-loss single-channel SS-OCT system: (a) inverted taper for efficient bulk/fiber optic component coupling to and from the PIC; (b) multimode interferometer (MMI) optimized for TM-like polarization to interfere incoming light from the sample and reference paths; (c) directional coupler to split the optical input path into sample and reference path with 90\% of the optical power on the through port and 10\% on the cross port; (d) polarization beam splitter for redirecting the returning light in the orthogonal polarization state towards the interferometer with TE-like polarization suppressed towards the cross port and vice versa for the TM-like polarization. 
}
\label{fig:component_summary}
\end{figure}

\begin{figure}[ht]
\centering
\includegraphics[width=13.3cm]{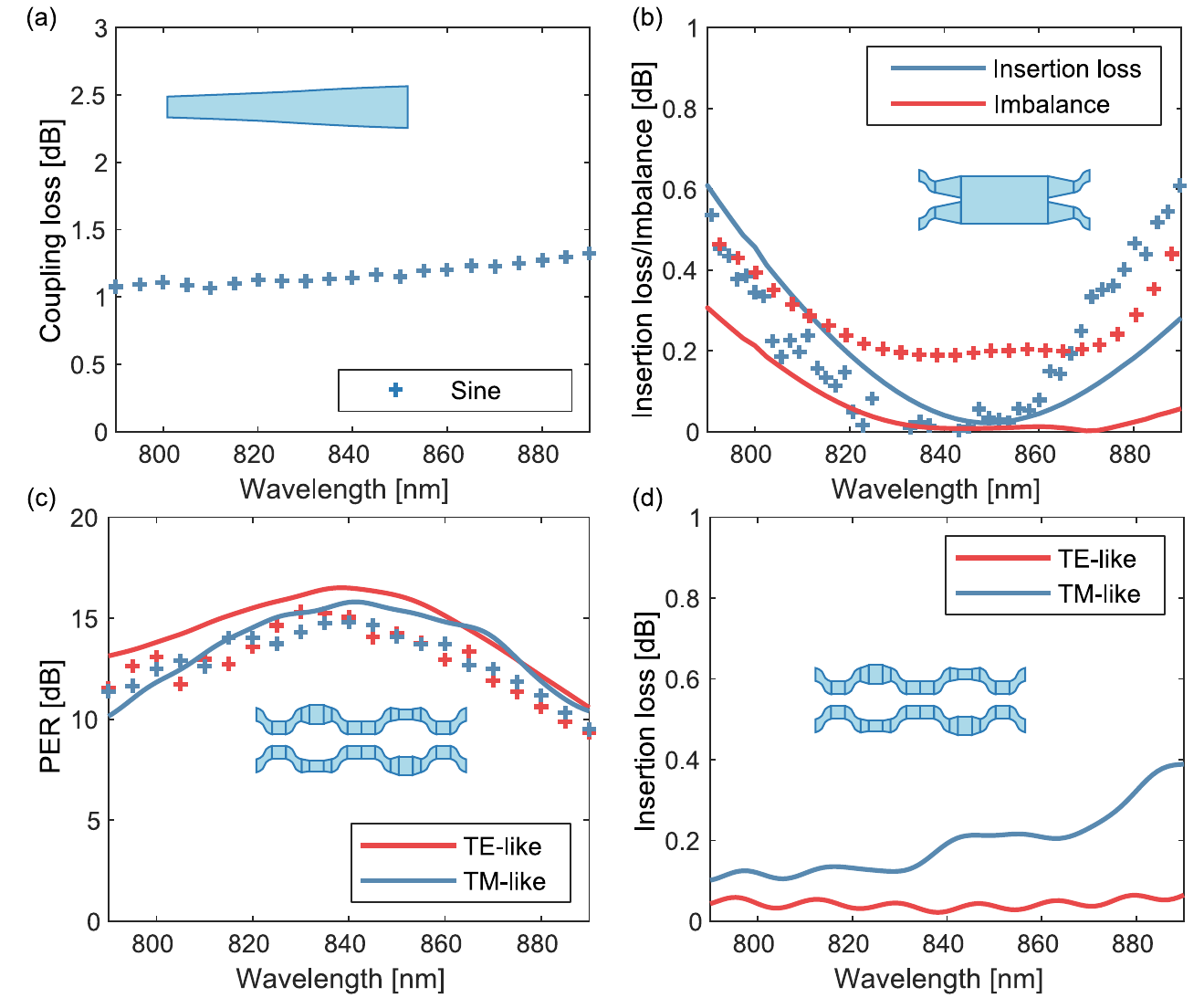}
\caption{Measurement (crosses) and simulation (lines) results of photonic building blocks: (a) insertion loss for the inverted taper with sinusoidal width envelope function (averaged from three different samples); (b) insertion loss and imbalance of the MMI, both determined with a symmetric Mach-Zehnder interferometer comprising two MMIs; (c) polarization extinction ratio (PER), and (d) losses of the PBS for both polarizations. The simulations for the MMIs were made using the eigenmode expansion (EME) method in FIMMPROP (Photon Design), whereas the simulations for the PBS were performed with the 3D FDTD method.}
\label{fig:component_measurement}
\end{figure}

There are four essential photonic building blocks required for the proof-of-concept single-channel PIC-based SS-OCT system: a mode size converter to couple light efficiently to and from the PIC, a power splitter to split the optical power into reference and sample path, a polarization beam splitter to redirect the light in the optical return path, and finally an interferometer. To avoid degradation of the axial resolution in an OCT measurement all of these photonic building blocks have to exhibit broadband wavelength behaviour. The design bandwidth was set to \SI{100}{\nano\metre}. For the mode size conversion an inverted taper was designed, which is depicted in Fig.~\ref{fig:component_summary}(a). In this photonic building block the waveguide width is reduced to lower the confinement and expand the mode into the cladding material. At the minimum width of the taper the resulting mode field diameter and the corresponding numerical aperture provide a better match with fiber and bulk optic components resulting in lower coupling losses. Due to the underlying operation principle inverted tapers have a low wavelength dependence and, thus, are well suited for broadband wavelength applications. To ensure a smooth mode transition at the beginning and the end of the taper a sinusoidal envelope function (covering a half period) is used for the width. Figure~\ref{fig:component_measurement}(a) shows the measurement results for the inverted taper using a polarization maintaining high NA fiber at the input in combination with index matching fluid. The coupling loss amounted to \textasciitilde 1.2 dB.
The design (see Fig.~\ref{fig:component_summary}(c)) and performance of the broadband 90/10 directional coupler for TE-like polarization at \SI{840}{\nano \metre}, acting as power splitter, has been reported in a previous publication \cite{Nevlacsil.2018}. The broadband behaviour is achieved by introducing a phase shift section between two directional couplers. The broadband polarization beam splitter uses the same phase shifting concept as the broadband directional coupler with an additional point symmetry \cite{Lu.2015}, as shown in Fig.~\ref{fig:component_summary}(d). Figure~\ref{fig:component_measurement}(c) shows the measurement results in terms of polarization extinction ratio defined as the ratio of optical power in the through port to the power in the suppressed port. Photonic building blocks based on directional couplers exhibit comparatively low insertion losses since they do not have abrupt changes in the cross section, which makes an experimental evaluation of the insertion loss difficult. Hence, the insertion losses (see Fig.~\ref{fig:component_measurement}(d)) have only been determined with 3D finite difference time domain (FDTD) using a commercial tool from Lumerical. The final photonic building block required to build a single-channel SS-OCT system is the interferometer. For this purpose, a $2\times2$ multimode interferometer (MMI) was designed (see Fig.~\ref{fig:component_summary}(b)). MMIs are especially well suited for interference because they can be designed with a low power imbalance across a wide wavelength range, which is of utmost importance for balanced detection. Furthermore, they feature a high fabrication tolerance because of their relatively large dimension compared to the width variations caused by the lithography process. Figure~\ref{fig:component_measurement}(b) shows the measurement results indicating insertion losses \textcolor{black}{(IL)} below \SI{0.7}{dB} and a \textcolor{black}{power} imbalance \textcolor{black}{(IB)} below \textcolor{black}{\SI{0.5}{dB}} in the targeted wavelength range. \textcolor{black}{These MMI results were experimentally determined with a symmetric Mach-Zehnder interferometer (MZI) consisting of two cascaded MMIs in a back-to-back configuration \cite{Halir.2009}. The insertion loss is defined by
\begin{equation}
\text{IL} [\text{dB}] = \frac{-10\,\text{log}_{10}\,(P_1+P_2)}{2},
\end{equation}
where $P_{1,2}$ are the normalized optical powers measured at the MZI output ports. The factor 2 results from the cascaded MMI configuration. The imbalance, which is defined as the ratio between the output powers of the two MMI output ports, was derived from the measured extinction ratio (ER) using the following relationship
\begin{equation}
\begin{split}
\text{ER} [\text{dB}] & = 10\,\text{log}_{10}\,(P_{1}/P_{2}) \\
&= 10\,\text{log}_{10}\,\bigg(\frac{4}{\rho+\rho^{-1}-2\text{cos}(2\theta)}\bigg),
\end{split}
\end{equation}
where $\rho = 10{\mathchar"5E}(\text{IB}/10)$ is the imbalance in linear scale and $\theta$ is the phase error. Assuming $\theta=0$ the resulting maximum power imbalance is 
\begin{equation}
\text{IB}_{\text{max}} [\text{dB}] = 10\,\text{log}_{10}\,\bigg(1+\frac{2}{\eta}\Big(1+\sqrt{1+\eta}\Big)\bigg),
\end{equation}
where $\eta = 10{\mathchar"5E}(\text{ER}/10)$ is the extinction ratio in linear scale.}

\begin{figure}[ht]
\centering
\includegraphics[width=13.3cm]{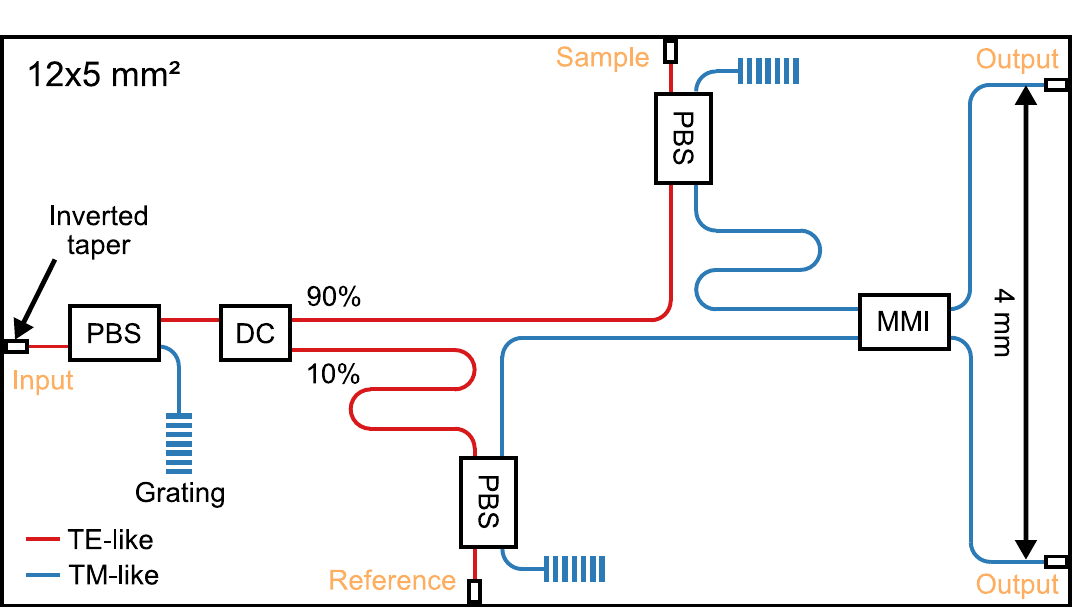}
\caption{\textcolor{black}{Schematic of PIC for single-channel SS-OCT system}: The input laser light is coupled to the TE-like mode on the PIC from the west side via the inverted taper. The grating coupler located at the suppression port of the PBS provides optical feedback for adjusting an external polarization controller to optimize the input polarization. The TE-like polarized light is then separated by the directional coupler with 90\% of the light going into the sample path (north) and 10\% going into the reference path (south). Both in the sample and the reference path a PBS redirects the returning polarization rotated light towards the MMI. The interfered light exits the output ports on the east side, which are spatially separated to avoid cross talk between the output ports. The length of the sample and the reference paths are matched to avoid wavelength dispersion mismatch on the PIC.}
\label{fig:1_channel_OCT_setup}
\end{figure}

Figure~\ref{fig:1_channel_OCT_setup} depicts the \textcolor{black}{schematic of the PIC that is used in a} single-channel SS-OCT system comprising the presented photonic building blocks. The lengths of the reference and sample paths on the PIC are identical for each polarization to minimize dispersion mismatch.

\begin{figure}[ht]
\centering
\includegraphics[width=10cm]{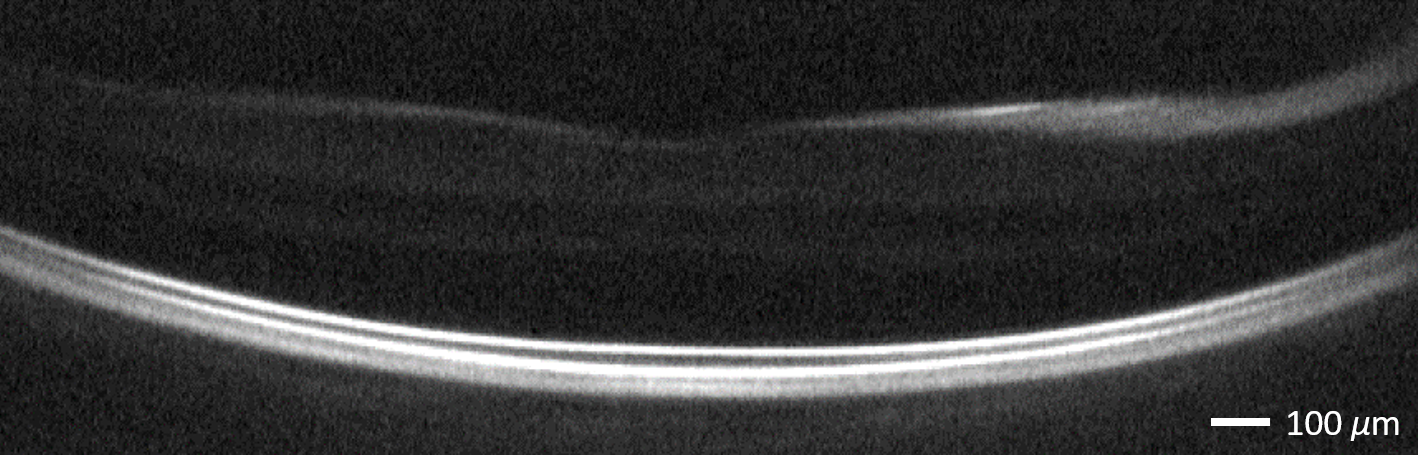}
\caption{Tomogram \textcolor{black}{(averaged over 400 consecutive B-scans)} of a healthy human retina acquired with the PIC-based single-channel SS-OCT system. The layer structure of the retina can be clearly seen. \textcolor{black}{The axial resolution appears to be lower than the determined value of \SI{5.5}{\micro\metre} because of the high number of averages resulting in a blurring of the image due to insufficient motion correction of the micro movements of the eye ball.}}
\label{fig:OCT_measurement}
\end{figure}

\textcolor{black}{Preliminary} OCT measurements with the fabricated PIC were conducted at the Medical University of Vienna. A \SI{840}{\nano\metre} swept laser source with a bandwidth of \SI{60}{\nano \metre} was provided by EXALOS. Figure~\ref{fig:OCT_measurement} shows the tomogram of a healthy human retina measured with the fabricated PIC. This preliminary measurement demonstrates that in-vivo tomograms of the retina layer structure can be achieved with a PIC-based SS-OCT system using the proposed routing approach. \textcolor{black}{Some trade-offs were made to allow the use of available fiber/bulk components, which reduced the overall system performance independent of the PIC. Instead of the $\lambda/4$-plates in the sample and reference paths fiber-based paddle polarization controllers were used to manipulate the polarization. This results in a reduced broadband capability of the system compared to using an achromatic $\lambda/4$-plate. Furthermore, the polarization controllers use standard single mode fibers and therefore the high NA fiber had to be replaced, which resulted in a $\approx$\,\SI{1.5}{dB} penalty compared to the best achievable value for the coupling loss. This was the case at the input, sample, and reference ports.} \textcolor{black}{The} interfered light exiting the PIC end facet was projected onto the photodiodes of a commercially available dual balanced detector \textcolor{black}{in free-space} using multiple lenses and mirrors. This induced high excess loss, which resulted in lower contrast in the tomogram. An optimized geometric configuration of the dual balanced photodiodes matching the distance of the output ports is expected to significantly improve the efficiency. \textcolor{black}{In this system a sensitivity of \SI{79}{dB} was measured with a sample beam power of \SI{0.6}{mW} and an acquisition rate per sample position (A-scan) of \SI{20}{kHz}. The axial resolution was determined as \SI{5.5}{\micro\metre} in air from the FWHM of the signal peak generated with a mirror as sample. This value is close to the theoretical limit of \SI{5.2}{\micro\metre} for a central wavelength of \SI{840}{\nano\metre} and a bandwidth of \SI{60}{\nano\metre}. This result indicates that the PIC provides good broadband performance in the investigated wavelength region.}

\section{Performance analysis}

With the knowledge of the optical characteristics of the individual building blocks the potentially achievable performance of a PIC-based multi-channel SS-OCT system can be estimated. Although parallelization brings an obvious advantage in the imaging speed, there will be no net gain in the performance of a multi-channel system if the same imaging quality as that of a single-channel system can only be obtained by averaging over multiple scans with the number of scans being larger than the number of parallel channels. As benchmark for this analysis we consider an ideal lossless single-channel system with an A-scan rate of \SI{100}{kHz}. In our analysis, we assume that the sweep rate of the swept source can be reduced arbitrarily to compensate losses introduced by the photonic building blocks for achieving the same SNR as the lossless system. We define the effective A-scan rate as the adjusted source sweep rate multiplied by the number of channels and use it as key performance indicator. In our concept the returning light from the eye experiences the same losses for each OCT channel independent of the number of channels. Consequently, the SNR is not reduced by increasing the number of channels. Therefore, the multi-channel effective A-scan rate is simply the A-scan rate of the lossy single-channel system multiplied by the number of channels. The only limitation stems from the available source power, which ultimately defines the maximum number of OCT channels. By increasing the number of OCT channels further losses are introduced requiring more source power to reach the desired single spot optical power on the sample. To determine the required optical source power the losses in the forward optical path towards the eye have to be calculated. 

We consider a multi-channel SS-OCT PIC with $2^N$ output ports, where $N$ is the number of cascading steps. As shown in Fig.~\ref{fig:4_channel_PIC}, in a multi-channel system there are two additional photonic building blocks compared to the single-channel configuration discussed in the previous section -- broadband Y-branches for equal power splitting to the individual channels and dual polarization crossings. We have already implemented both building blocks for the \SI{840}{\nano\metre} wavelength region \cite{Hainberger.01.04.2019,Nevlacsil.2020}. Table~\ref{tab:component_losses} provides a summary of the loss values of all building blocks in the multi-channel system for a wavelength of \SI{840}{\nano\metre}. Bend losses are not listed explicitly because radii can be increased at multiple positions making the accumulated propagation loss the main contribution. Although the losses are dependent on the wavelength, this primarily affects the achievable axial resolution rather than the SNR of the OCT system.

\begin{table}[ht]
\caption{Summary of the propagation loss (PL), insertion loss ($\Gamma$), and crosstalk (CT) values for the individual building blocks to be used in the proposed multi-channel SS-OCT system. All values are given in \si{dB} for the central wavelength of \SI{840}{\nano \metre}. They have been experimentally determined with fabricated devices except the ones indicated with $^S$, which have been derived from 3D FDTD simulations.}
\centering
\begin{tabular}{lll}
\hline
 & TE-like & TM-like \\ \hline
PL & 0.75/cm & 0.35/cm \\
$\Gamma_{\text{Inverted taper}}$ & 1.14 & 1.14\\
$\Gamma_{\text{DC}}^S$ & 0.02& -\\
$\Gamma_{\text{PBS}}^S$ & 0.02 & 0.19\\
$\text{CT}_{\text{PBS}}$ & 0.14 & 0.15\\
$\Gamma_{\text{Crossing}}$ & 0.07 & 0.02\\
$\Gamma_{\text{Y-branch}}$ & 0.11 & 0.14\\
$\Gamma_{\text{MMI}}$ & - & 0.01\\ \hline
\end{tabular}
\label{tab:component_losses}
\end{table}

In reference to Fig.~\ref{fig:4_channel_PIC}, for the loss of source power towards the eye two types of losses can be distinguished -- losses that scale with the number of cascading steps and losses that do not. The number of Y-branches and crossings along the optical path of one channel equals the number of cascading steps $N$. For the Y-branches both the insertion losses and the equal splitting (\SI{3}{dB}) have to be included. Concerning the propagation length scaling with the number of cascading steps, the additional length in x-direction is given by the spacing of the channels. The channel spacing is set to \SI{500}{\micro \metre}, which is a typical value for microlens arrays. In y-direction, the additional length is determined by the length of the Y-branches, the crossings and the bends as shown in Fig.\ref{fig:4_channel_PIC}. In total, this results in an additional path length of approximately $N \times \SI{850}{\micro \metre}$. The loss contributions independent of the number of channels are dominated by the single-channel photonic building blocks because the propagation lengths of most sections can be minimized by adapting the PIC dimension. The only exception is the section from the cascading tree up to the sample ports measuring about \SI{700}{\micro \metre} due to the MMI length and bends from the MMI to the PDs, with the PD spacing being half of the channel spacing. The losses of photonic building blocks that have to be included are from the DC, both in terms of insertion loss and amount of light directed to the reference path (only\SI{-0.46}{dB} or $\approx$90\% towards sample), the two PBS, in terms of insertion and crosstalk loss, and the two inverted tapers at the input and sample port. In summary, the power is reduced from the source to the sample by
\begin{equation}
\begin{split}
\Gamma_{\text{Source}}[\si{dB}] & = 2\Gamma_{\text{Inverted taper}}+\Gamma_{\text{DC}}+\SI{0.46}{dB}+2(\Gamma_{\text{PBS}}+\text{CT}_{\text{PBS}})+\SI{700}{\micro\metre}\times \text{PL}\\
& + N\times(\Gamma_{\text{Y-branch}}+\SI{3}{dB}+\Gamma_{\text{Crossing}}+\SI{850}{\micro \metre}\times \text{PL}) \\
& \approx \SI{3.313}{dB} + N \times \SI{3.244}{dB}\textcolor{black}{.}
\end{split}
\end{equation}
The loss $\Gamma_{\text{Source}}$ results from light propagating in TE-like polarization. For the loss of signal light three photonic building blocks are relevant -- the inverted taper, PBS and MMI. Since the components can be connected directly to each other only the connection bends from the MMI to the photodiodes are relevant, which are approximately \SI{350}{\micro \metre} in length. In conclusion, the loss from the sample to the photodiode is calculated with
\begin{equation}
\Gamma_{\text{Signal}}[\si{dB}] = \Gamma_{\text{Inverted taper}}+ \Gamma_{\text{PBS}}+\text{CT}_{\text{PBS}}+\Gamma_{\text{MMI}}+\SI{350}{\micro\metre}\times \text{PL} \approx \SI{1.502}{dB},
\end{equation}
from light propagating in TM-like polarization.
\newpage
For the calculation of the required source power the intended power at the sample is divided by the power reduction $\Gamma_{\text{Source}}$. We assume a single beam power of \textcolor{black}{\SI{0.5}{\milli \watt}} to guarantee laser safe operation at \SI{840}{\nano\metre} according to the \textcolor{black}{norms ISO 15004 and ANSI Z80.36.. In the case that the single beam is scanned over the retina a higher power of \SI{0.75}{\milli\watt} would be allowed resulting in an average power of less than \SI{0.5}{\milli\watt}. While the higher allowed power was used in the single-channel measurement, for a multi-channel system a more restrictive consideration in regard to safety is recommendable.} \textcolor{black}{For the safety of a multi-beam ophthalmic system the combined effect of the beams has to be taken into account. While spatially separated beams at the retina can be considered independently, at the cornea and lenticle the beams may overlap and the power may add up. The norms give a safe mean intensity of \SI{100}{mW/cm\squared} averaged over a defined area at the corneal plane. According to the ANSI norm more than sixteen single beams with 0.5 mW each can safely overlap at the cornea.} Table~\ref{tab:performance_analysis} shows the values for the required source power to reach \textcolor{black}{\SI{0.5}{\milli \watt}} at each probing position and the effective A-scan rate for an increasing number $N$ of cascading steps (channels).

\begin{table}[ht]
\caption{Performance analysis of multi-channel configurations taking into account losses of the designed building blocks for the center wavelength of \SI{840}{\nano \metre}. The first column shows the required source power to achieve an optical power of \textcolor{black}{\SI{0.5}{\milli \watt}} for each probing beam at the retina. The second column shows the resulting effective A-scan rate. \textcolor{black}{The SNR for all systems is identical to the lossless single-channel system because the sweeping rate is adapted accordingly for the lossy single-channel system and no additional loss is introduced on the way from the sample to the interferometer for higher number of channels.}}
\centering
\begin{tabular}{lll}
\hline
 & \begin{tabular}[c]{@{}l@{}}Source power \\ {[}mW{]}\end{tabular} & \begin{tabular}[c]{@{}l@{}}Effective A-scan rate \\ {[}kHz{]}\end{tabular} \\ \hline
1-channel & \textcolor{black}{1.1} & 71\\
2-channel & \textcolor{black}{2.3} & 142\\
4-channel & \textcolor{black}{4.8} & 284\\
8-channel & \textcolor{black}{10.1} & 568\\
16-channel & \textcolor{black}{21.3} & 1136\\
32-channel & \textcolor{black}{44.9} & 2272\\ \hline
\end{tabular}
\label{tab:performance_analysis}
\end{table}
As can be seen in the table, already for two channels the effective A-scan rate exceeds the \SI{100}{\kilo \hertz} figure of merit. For up to \textcolor{black}{32} channels the laser power stays below \SI{50}{\milli \watt}, which is realistically achievable for a swept source with a booster amplifier. The effective A-scan rate shows that the imaging can be done \textcolor{black}{with a sixteen-channel OCT system} at approximately ten times the speed of a lossless single-channel system with the same SNR. Therefore, even in the comparison with the unrealistic case of a lossless system the multi-channel system can achieve higher acquisition rates. 

\section{Outlook}
The next step towards a multi-channel system is to combine the designed building blocks with optoelectronics and electronics, e.g.\@ photodiodes and analog to digital converter (ADC), on the same chip. The inclusion of the reference path in the PIC, with the potential capability to change the length, would be a further step towards integration. However, this would require significantly lower propagation losses and dispersion. The dispersion issue can be addressed by changing the dispersion properties e.g.\@ by changing the cross section and/or the cladding material \cite{Yurtsever.2011}. To facilitate the numerical dispersion compensation a PIC-based k-clock can be included for accurate real-time information on the dispersion properties, consisting of an asymmetric Mach-Zehnder interferometer with a well defined physical path length difference.
\section{Conclusion}
In conclusion, we have presented a concept for a low-loss PIC-based multi-channel SS-OCT system utilizing the intrinsic advantages of PICs such as high integration density and mechanical stability enabling parallelization. In comparison, bulk or fiber optic multi-channel implementations are significantly more intricate to achieve with growing difficulties and losses for each additional measurement channel. As a proof of concept of the polarization based optical path routing concept, which forms the basis for our low-loss multi-channel concept, we designed and fabricated a \textcolor{black}{PIC for a single-channel SS-OCT system operating at} \SI{840}{\nano\metre} with a bandwidth of \SI{100}{\nano \metre} including all major photonic building blocks. We showed that the performance of this PIC-based OCT system is sufficient to achieve in-vivo retinal OCT tomograms. 
This work represents the first PIC-based SS-OCT demonstration in the \SI{840}{\nano\metre} wavelength domain. With the designed and tested photonic building blocks we carried out a performance analysis of multi-channel configurations. We showed that in principle the laser source power levels can be kept reasonable, while achieving a tenfold speed increase compared to an ideal single-channel system with the same SNR. This can be done even though excess losses in PICs are higher than in bulk and fiber optics. This shows that PICs have the potential to play an important role in future OCT development, not only in regard to size and cost but also to performance.
\section*{Funding}
H2020 LEIT Information and Communication Technologies (688173, 871312).
\section*{Acknowledgments}
This research has received funding from the European Union's Horizon 2020 research and innovation program under grant agreement No 688173 (OCTCHIP) and No 871312 (HandheldOCT). Portions of this work were presented at the SPIE OPTO in 2019 ''Silicon-nitride waveguide-based integrated photonic circuits for medical diagnostic and other sensing applications'', the SPIE Optics + Optoelectronincs in 2019 ''PECVD silicon nitride optical waveguide devices for sensing applications in the visible and <\SI{1}{\micro\metre} near infrared wavelength region'' and the ECIO in 2020 ''PECVD SiN photonic integrated circuit for swept source OCT at 840 nm''. \textcolor{black}{The authors want to thank Dr.\@ Michael Kempe and Daniel Bublitz of Carl Zeiss AG for their input on the laser safety norms.}
\section*{Disclosure}
The authors declare no conflicts of interest.
%

\end{document}